\documentclass[aps, twocolumn, showpacs, amsmath,superscriptaddress]{revtex4}
\usepackage{dcolumn}
\usepackage{bm}
\usepackage{graphicx}
\usepackage{color}
\usepackage{xcolor}
\DeclareGraphicsExtensions{. jpg,. pdf, . mps, . png, . eps, . ps, . EPS}

\def\bc{\begin{center}}
\def\ec{\end{center}}
\def\bea{\begin{eqnarray}}
\def\eea{\end{eqnarray}}
\newcommand{\avg}[1]{\langle{#1}\rangle}

\begin{document}

\title{Entropy distribution and condensation in random networks \\ with a given degree distribution}


\author{Kartik Anand}
\affiliation{Bank of Canada, 234 Laurier Ave W., Ottawa, Ontario K1A 0G9, Canada}
\author{Dmitri Krioukov}
\affiliation{Department of Physics, Northeastern University, Boston, MA 02115, USA}
\author{Ginestra Bianconi}
\affiliation{School of Mathematical Sciences, Queen Mary University of London, London, E1 4NS, UK}

\pacs{89.75.Hc,89.75.Da,64.60.aq}

\begin{abstract}
The entropy of network ensembles characterizes the amount of information encoded in the network structure, and can be used to quantify network complexity, and the relevance of given structural properties observed in real network datasets with respect to a random hypothesis. In many real networks the degrees of individual nodes are not fixed but change in time, while their statistical properties, such as the degree distribution, are preserved. Here we characterize the distribution of entropy of random networks with given degree sequences, where each degree sequence is drawn randomly from a given degree distribution. We show that the leading term of the entropy of scale-free network ensembles depends only on the network size and average degree, and that entropy is self-averaging, meaning that its relative variance vanishes in the thermodynamic limit. We also characterize large fluctuations of entropy that are fully determined by the average degree in the network. Finally, above a certain threshold, large fluctuations of the average degree in the ensemble can lead to condensation, meaning that a single node in a network of size~$N$ can attract $O(N)$ links.
\end{abstract}

\maketitle

\section{Introduction}

One reason why network science has recently attracted significant research attention is that network structure efficiently encodes the complexity of a large variety of systems, from the brain to different techno-social infrastructures \cite{Barabasi,Newman_book,Dorogovtsev_book,dyn}. In the last fifteen years or so there has been significant progress in characterizing not only universal properties of complex networks, e.g., scale-free degree distributions or small-world properties, but also their specific features that distinguish one network from another---degree correlations,  community structure \cite{Santo}, or motif distributions \cite{Alon,Loops,Cliques,massimo}.

More recently, considerable effort has focused on quantifying network complexity using new network entropy measures  \cite{Entropy,AB2009,BC2008,AB2010,PNAS,Munoz,Peixoto_BM,Peixoto_PRL,DG}  borrowed from information theory, statistical mechanics \cite{Newman1,Garlaschelli1,Garlaschelli2,Coolen2}, and quantum information \cite{AB2011,Garnerone}. The  entropy  of a  network ensemble  evaluates the total number of networks belonging to the ensemble \cite{Entropy,AB2009}. The more complex, sophisticated, and unique the network structure, the smaller the number of networks in the ensemble having these peculiar properties, the smaller the entropy. The entropy measures have proven useful for solving inference problems involving real-world networks \cite{PNAS,Munoz,Peixoto_BM,Peixoto_PRL}. The statistical mechanics treatment of network ensembles can be used to characterize the likelihood that a real dataset is generated  by a model \cite{Garl_l}.  Some real networks have been shown to belong with high likelihood to ensembles of random geometric graphs in hyperbolic spaces, modelling trade-offs between popularity and similarity in network evolution, and casting preferential attachment as an emergent phenomenon \cite{Krioukov1,Krioukov2}. More recently, the entropy of multiplex ensembles has been proposed to characterize the complexity of multilayer networks \cite{Multiplex}.

By definition, a network ensemble is a set of graphs $G$ with probability measure $P(G)$. It is important to make a distinction between microcanonical and canonical ensembles \cite{AB2009}. In microcanonical ensembles, some structural network properties are fixed to given values. For example, the total number of links in graphs of size $N$ can be fixed to $M$, or the degrees of all nodes can be fixed to degree sequence $\{k_i\}$, $i=1,\ldots,N$. In this case, the ensemble consists of all graphs of size $N$, and the probability measure is uniform: if the number of graphs $G$ satisfying the constraints is ${\mathcal N}$, then for all such graphs, $P(G)=1/{\mathcal N}$, and $P(G)=0$ for all other graphs that do not satisfy the constraints. In the canonical counterparts of these ensembles, the same structural constraints are fixed only on average---the resulting ensembles are maximum-entropy ensembles under the constraints that the expected values of the number of edges or node degrees in the ensemble are $M$ or $\{k_i\}$. The probability measure $P(G)$ in this case is not uniform---the closer the $G$ to satisfying the constraints, the larger the $P(G)$. In random graph ensembles with a fixed exact or expected number of links, the canonical distribution converges to the microcanonical distribution in the thermodynamic limit $N\to\infty$. However, as soon as the number of imposed constraints is extensive, the canonical and microcanonical network ensembles are not equal even in the thermodynamic limit, and neither are their entropies. For example, the entropy of the microcanonical ensemble with a fixed degree sequence is not equal to the entropy of the canonical ensemble where only the expected degree sequence is fixed \cite{AB2009}.

In network theory there is usually no question of how precisely we know the node degrees: given a graph, its degree sequence is uniquely defined. However when a network practitioner works with real network data, she is typically given a collection of network measurements. Does she have to treat the measured degrees of nodes as precisely defined as well, given that measurements are always imprecise and ever-changing, and so is the network itself? The answer is usually `no'---the relevant information is not the exact degree sequence, but its statistical properties, such as the distribution of these degrees. The ensembles of networks with a given exact or even expected degree sequence do not account for possible statistical fluctuations of node degrees in a given dataset, motivating us to consider here ensembles of random networks whose exact or expected degree sequences $\{k_i\}$ are independently sampled from a given distribution $p(k)$. This approach is a way to explore only the statistical properties of networks, and not their specific linking diagrams that  might be affected by false or missing links, almost always present in real data.

Specifically, we study the distribution of entropy in network ensembles with a given exact or expected degree distribution $p(k)$. We find that in both cases (hard/exact and soft/expected), if the network ensemble is sparse, the average entropy is well-defined and self-averaging.
We also evaluate the probability that ensemble entropy is equal to a particular value, conditioned on the total number of links in the network, and show that this conditional entropy distribution is always well-behaved.
Characterizing  large entropy deviations in the ensemble with a given degree distribution and average degree, we observe that a condensation phenomena can occur in the network.
This phenomena occurs only if the average degree in the network $\avg{k}$ exceeds the degree distribution average $m=\sum_k kp(k)$. For $\avg{k}<m$ there is a symmetry under permutation of the labels of the nodes of the network, meaning that if we fix the average degree $\avg{k}$ of the network, then the degrees of all nodes are $o(N)$, where $N$ is the network size. Instead, if $\avg{k}>m$, we observe a spontaneous breaking of this symmetry, with a single node having an $O(N)$ degree. These results hold in both hard and soft ensembles, i.e., ensembles with fixed exact or expected degree distributions.

We begin with reviewing in Section~II what is known about the entropy of network ensembles with a fixed expected or exact degree sequence. We then move to  Section~III and~IV where we analyze some properties of the entropy distributions in the ensemble with a given  expected and exact degree distribution, respectively. Final remarks are in Section~V.

\section{Entropy of network ensembles with a given degree sequence}
A network ensemble is specified once probability $P(G)$ is assigned to every network $G$ of size $N$. We denote nodes by $i=1,2,\ldots,N$. The set of simple undirected unweighted labeled networks of size $N$ is bijective to the set of symmetric boolean $N\times N$ adjacency matrices ${\bf a}\in\{0,1\}^{N\times N}$ having zeroes on the diagonal. Depending on whether nodes $i$ and $j$ are connected or not in network $G$, element $a_{ij}$ of $G$'s adjacency matrix is either $1$ or $0$.
We next impose the constraint that the degree $\sum_ja_{ij}$ of each node $i$ is fixed to some $k_i$. We can treat this constraint as hard or soft. If it is hard,
we impose it exactly. The resulting ensemble is a microcanonical network ensemble with given degree sequence $\{k_i\}$, known as the configuration model \cite{MR,Chung}.  In the soft case, we relax the constraint, and demand that the degree of each node $i$, averaged over all networks in the ensemble, is $k_i$, which no longer has to be integer but can be any non-negative real number. The resulting ensemble is a canonical network ensemble with a given expected degree sequence $\{k_i\}$, belonging to the class of random graphs known as exponential random graphs \cite{Caldarelli, Boguna, Newman1,dyn}.  In what follows we denote by $P_e(G|\{k_i\})$ the probability of $G$ in the canonical ($e=C$) or micro-canonical ($e=M$) ensembles.
The entropy $S(\{k_i\})$ of these  ensemble evaluates the typical number of networks in the ensemble and is given by
\bea
S=-\sum_{G}P_e(G|\{k_i\})\ln P_e(G|\{k_i\})\,,
\label{entropy0}
\eea
where the sum is performed over all networks in the ensemble.

\subsection{Entropy of the canonical ensemble}

The probability distribution $P_C(G|\{k_i\})$ in the canonical ensemble is defined as the distribution that maximizes entropy
\bea
S(\{k_i\})=-\sum_{G}P_C(G|\{k_i\})\ln P_C(G|\{k_i\})\,,
\label{entropy}
\eea
subject to the following $N$ constraints:
\begin{equation}
\sum_{G}P_C(G|\{k_i\})\sum_{j\ne i}a_{ij}=k_i,\quad{\rm for}\quad i=1,\ldots,N\,.
\label{constraints}
\end{equation}
By summing over all networks $G$, we sum over their adjacency matrices. Introducing Lagrangian multipliers $\lambda_{i}$ to enforce the conditions in Eq.~$(\ref{constraints})$, and Lagrangian multiplier $\Lambda$ to normalize the probability measure $\sum_{G}P_C(G|\{k_i\})=1$, we solve the system of equations
\begin{eqnarray}
\frac{\partial }{\partial P(G| \{k_i\})}&\left[S-\sum_{i=1}^N \lambda_{i} \sum_{{G}}\sum_{j\neq i} a_{ij}P_C(G|\{k_i\})\nonumber \right.\\
&\left.-\Lambda\sum_{G}P(G|\{k_i\})\right]=0
\end{eqnarray}
to find the expression for distribution $P_C(G|\{k_i\})$:
\begin{equation}
P_C(G|\{k_i\})=\frac{1}{Z_C}\exp\left[-\sum_{i=1}^N \sum_{j\neq i}\lambda_i a_{ij},\right]
\label{PC}
\end{equation}
where the normalization constant
\bea
Z_C=e^{-\Lambda}=\sum_{G}\exp\left[-\sum_{i=1}^N \sum_{j\neq i}\lambda_i a_{ij}\right]
\eea
is called the  ``partition function.''
Since the probability distribution in Eq.~$(\ref{PC})$ has an exponential form, this ensemble is called {\em exponential random graphs}.

In this ensemble, we can  relate entropy $S(\{k_i\})$ in Eq.~$(\ref{entropy})$  to partition function $Z_C$:
\begin{eqnarray}
S(\{k_i\})&=&-\sum_{{G}} P_C({G}|\{k_i\}) \ln  P_C({G}|\{k_i\})\nonumber \\&=&-\sum_{{G}} P_C({G}|\{k_i\})\left[-\sum_{i=1}^N \lambda_{i} \sum_{j\neq i} a_{ij}-\log(Z_C)\right]\nonumber \\
&=&\sum_{i=1}^N \lambda_{i} k_i+\log{Z_C}.
\label{Sc}
\end{eqnarray}
We call the entropy $S(\{k_i\})$ of the canonical ensemble the {\it Shannon entropy}.

The probability $p_{ij}$ of a link between node $i$ and node $j$ in the ensemble is given by
\bea
p_{ij}=\avg{a_{ij}}&=&\frac{e^{-(\lambda_i+\lambda_j)}}{1+e^{-(\lambda_i+\lambda_j)}}\nonumber \\
&=&\frac{h_ih_j/N}{1+h_ih_j/N},\label{pijc}\\
\eea
where $h_i=\sqrt{N}e^{-\lambda_i}$
are called ``hidden variables''. Upon this change of variables, the constraints in  Eq.~$(\ref{constraints})$ translate to
\bea
k_i=\sum_{j\neq i}p_{ij}=\sum_{j\neq i}\frac{h_ih_j/N}{1+h_ih_j/N}.
\label{kpij}
\eea
This system of equations can be solved for $\{h_i\}$ yielding the values of Lagrangian multipliers $\{\lambda_{i}\}$.
Using $p_{ij}$ in Eq.~$(\ref{pijc})$, a simpler expression for distribution $P_C(G|\{k_i\})$ reads
\bea
P_C(G|\{k_i\})=\prod_{ij}p_{ij}^{a_{ij}}(1-p_{ij})^{1-a_{ij}}.
\label{eq:P_C(pij)}
\eea
Therefore probability $P_C(G|\{k_i\})$ is actually the probability to generate network $G$ with hidden variables $\{h_i\}$ by connecting node pairs $i$ and $j$ with probability $p_{ij}$ given by Eq.~(\ref{pijc}), and not connecting them with probability $1-p_{ij}$.
Using these link existence probabilities $p_{ij}$, the entropy of the ensemble in Eq. $(\ref{entropy})$ can be written as
\bea
S(\{k_i\})=-\sum_{i<j}[p_{ij}\log p_{ij}+(1-p_{ij})\log(1-p_{ij})].
\label{S}
\eea
By inserting the explicit dependence of probabilities $p_{ij}$ on hidden variables $h_i$, we can extract the leading term ${\cal S}(\{k_i\})$ of the entropy that depends only on the average degree in the network, and the subleading term $N\sigma(\{k_i\})$ that increases linearly with $N$:
\bea
S(\{k_i\})={\cal S}(\{k_i\})-N\sigma(\{k_i\}),
\label{S2}
\eea
where ${\cal S}(\{k_i\})$ and $\sigma(\{k_i\})$  in any sparse network are given by
\bea
{\cal S}(\{k_i\})&=&\frac{1}{2}\avg{k}N\ln N,\nonumber \\
N\sigma(\{k_i\})&=&\sum_{i<j}[p_{ij}\log (N p_{ij})+(1-p_{ij})\log(1-p_{ij})]\nonumber \\
&=&\left[\sum_{i<j}\frac{h_i h_j/N}{1+h_ih_j/N}\ln\left(\frac{h_i h_j}{1+h_ih_j/N}\right)+\right.\nonumber \\
&&\hspace*{-15mm}\left.\sum_{i<j}\left(1-\frac{h_i h_j/N}{1+h_ih_j/N}\right)\ln\left(1-\frac{h_i h_j/N}{1+h_ih_j/N}\right)\right].
\label{sigma}
\eea
If all expected degrees $k_i\ll\sqrt{\avg{k}N}$, where $\avg{k}$ is the average expected degree $\sum_ik_i/N$, then hidden variables $h_i\ll \sqrt{N}$ are proportional to expected degrees $k_i$, $h_i=k_i/\avg{k}$, and we can  approximate probabilities $p_{ij}$ in Eq.~$(\ref{pijc})$ by
\bea
p_{ij}=\frac{h_ih_j}{N}=\frac{k_i k_j }{\avg{k}N}\,.
\eea
which corresponds to the case where links in the network are uncorrelated.

In this case the expression for the extensive entropy term $\sigma(\{k_i\})$ in Eq.~$(\ref{sigma})$ simplifies to
\bea
\sigma(\{k_i\})&=&\frac{1}{N}\sum_i k_i \ln k_i-\frac{1}{2}\avg{k}[1+\ln{\avg{k}}].
\eea

\subsection{Entropy of the microcanonical ensemble}

In the microcanonical ensemble, all networks satisfy the hard constraint that the degree sequence is $\{k_i\}$ exactly.
We assume here that the degree sequence is graphical \cite{DelGenio1,DelGenio2}, meaning that it can be realized by at least one network. This condition is obviously satisfied if the degree sequence is read off from a real network.
The probability distribution in the ensemble is uniform---all networks satisfying this constraint have the same probability
\begin{equation}
P_M({G}|\{k_i\})=\frac{1}{Z_M}\prod_{i=1}^N \delta\left[\sum_{j\neq i}a_{ij}, k_i\right]
\label{PM}
\end{equation}
where $\delta[\ldots]$ stands for the Kronecker delta, and where ``partition function'' $Z_M$ is given by
\begin{equation}
Z_M=\sum_{{G}}\prod_{i=1}^N \delta\left[\sum_{j\neq i}a_{ij},k_i\right].
\label{ZM}
\end{equation}
This partition function simply counts the number of networks with degree sequence $\{k_i\}$.

The definition of the network ensemble entropy in Eq.~$(\ref{entropy0})$ applied to the microcanonical distribution in Eq.~$(\ref{PM})$ yields
\begin{equation}
N\Sigma(\{k_i\})=-\sum_{{G}}P_M({G})\ln P_M({G})=\ln Z_M,
\label{SM}
\end{equation}
where we call $N\Sigma(\{k_i\})$ the {\it Gibbs entropy} of the network ensemble.
The Gibbs entropy $N\Sigma(\{k_i\})$ of the microcanonical ensemble is related to the Shannon entropy $S(\{k_i\})$ of the conjugate canonical ensemble via
\begin{equation}
N\Sigma(\{k_i\})=S(\{k_i\})-N\Omega(\{k_i\}),
\label{SSO}
\end{equation}
where $N\Omega(\{k_i\})$ is  equal to the logarithm of the probability that in the conjugate canonical ensemble the hard constraints $\sum_{j\neq i}a_{ij}=k_i$ are satisfied:
\begin{equation}
N\Omega(\{k_i\})=-\log\left\{\sum_{{G}}P_C({G}|\{k_i\})\prod_{i=1}^N\delta\left[\sum_{j\neq i}a_{ij},k_i\right]\right\}.
\label{Omega}
\end{equation}
The relation between entropies in Eq.~$(\ref{SSO})$ can be obtained by substituting the canonical distribution $P_C({G}|\{k_i\})$ given by Eq.~$(\ref{PC})$ into Eq.~$(\ref{Omega})$, yielding
\begin{eqnarray}
\exp[-N\Omega(\{k_i\})]&=&\sum_{G} \frac{1}{Z_C}e^{-\sum_{i=1}^N\lambda_{i}\sum_{j\neq i}a_{ij}}\nonumber\\
&&\times\prod_{r=1}^N\delta\left[\sum_{s\neq r}a_{rs},k_r\right]\nonumber \\
&=&\frac{1}{Z_C}e^{-\sum_{i=1}^N \lambda_{i}k_i}\nonumber\\
&&\times\sum_{{G}}\prod_{r=1}^N\delta\left[\sum_{s\neq r}a_{rs},k_r\right]\nonumber \\
&=& \frac{Z_M}{e^{S(\{k_i\})}}=\exp[N\Sigma(\{k_i\})-S(\{k_i\})],\nonumber
\end{eqnarray}
where in the last relation we have used Eq.~$(\ref{Sc})$, Eq.~$(\ref{ZM})$, and Eq.~$(\ref{SM})$.
The value of function $\Omega(\{k_i\})$ in sparse networks can be calculated by statistical mechanics methods \cite{BC2008,AB2010}:
\bea
\Omega(\{k_i\})&=&\frac{1}{N}\sum_{i=1}^N \ln\left[\frac{k_i!}{\left(k_i/e\right)^{k_i}}\right].
\label{Oe}
\eea
It does not vanish in the thermodynamic limit $N\to\infty$. Therefore in view of the relation between the microcanonical and canonical entropies in Eq.~$(\ref{SSO})$, the microcanonical and conjugate canonical ensembles are not equivalent even in the large-$N$ limit.

\section{Entropy distribution in the network ensemble with a given distribution of expected degrees}

In this section we consider the network ensemble in which the expected degree sequence $\{k_i\}$ is not fixed but sampled in each network realization from a fixed distribution $p(k)$. Drawing from the field of disordered systems, we make a distinction between {\em quenched} and {\em annealed} disorder.
If the disorder is annealed the degree of the nodes are not fixed and they are continuously drawn form a degree distribution $p(k)$.
If the disorder is quenched, then the expected degree sequence $\{k_i\}$ in each realization is assumed to be fixed but unknown, and for each expected degree sequence, the ensemble probability distribution is obtained by maximizing the entropy. Below we consider the quenched case only.

For a fixed expected degree sequence $\{k_i\}$, the maximum-entropy distribution $P_C(G|\{k_i\})$ is given by Eq.~(\ref{eq:P_C(pij)}), while the entropy $S(\{k_i\})$ of this distribution is given by Eq.~(\ref{S}).
If this degree sequence $\{k_i\}$ has probability $P(\{k_i\})$ in a larger ensemble, then the probability and entropy distributions in this larger ensemble are
\bea
P_C(G)&=&\int {\prod_{i=1}^Nd{k_i}}  P(\{k_i\}) P_C(G|\{k_i\}),\\
P_C(S)&=&\int {\prod_{i=1}^Nd{k_i}}  P(\{k_i\}) \delta\left[S,S(\{k_i\})\right].
\eea
Therefore the   distribution  of the entropy $S(\{k_i\})$ in this ensemble gives a very important indication on how the number of possible network realization with expected degree sequence $\{k_i\}$ changes if the  sequence realization  is drawn randomly from a degree distribution $p(k)$.
If we cannot compute the full distribution $P_C(S)$ exactly, we may still characterize
its average, variance, and relative error
\bea
\overline{S(\{k_i\})}&=& \int {\prod_{i=1}^N dk_i}   P(\{k_i\}) S(\{k_i\})\nonumber \\
\overline{\left[\delta S(\{k_i\})\right]^2}&=& \int \prod_{i=1}^Nd{k_i}  P(\{k_i\})
 \left[S(\{k_i\})-\overline{S(\{k_i\})}\right]^2,\nonumber \\
 \Delta S(\{k_i\})&=&\frac{\sqrt{\overline{\left[\delta S(\{k_i\})\right]^2}}}{\overline{S(\{k_i\})}}.
\eea

\subsection{Entropy distribution in scale-free networks}

We assume that each expected degree sequence $\{k_i\}$ has probability $P(\{k_i\})=\prod_{i=1}^N p(k_i)$, where $p(k)$ is the expected degree distribution,
and consider the specific case of power-law $p(k)\simeq k^{-\gamma}$ with $\gamma\in (2,\infty)$.
We first focus on the leading term of entropy $S(\{k_i\})$ given by ${\cal S}(\{k_i\})=\frac{1}{2} \avg{k}N \ln N$, where   $\avg{k}N$ is the sum of expected degrees $\avg{k}N=\sum_{i=1}^N k_i$.

We distinguish between two cases.
\begin{itemize}
\item {\it Case $\gamma>3$.}\\
When $\gamma>3$, distribution $P({\cal S})$ is Gaussian,   the average of ${\cal S}$ is well defined in the network and its relative error vanishes in the large network limit. Indeed, since $\avg{k^2}<\infty$, we have
\bea
\Delta {\cal S}(\{k_i\})\propto N^{-1/2}.
\eea

\item{\it Case $\gamma\in (2,3]$.}\\
For large $N$ and $\gamma\in (2,3]$, due to the structural degree cutoff $k_i\leq k_{\max}=N$, we observe that   the average of ${\cal S}$ is also well defined in the network and its relative error also vanishes in the large network limit, but with a different exponent. Indeed,
since $\avg{k^2}\propto N^{3-\gamma}$ we have
\bea
\Delta {\cal S}(\{k_i\})\propto N^{-(\gamma-2)/2}.
\eea

\end{itemize}
These results are important because they imply  that for every value of $\gamma>2$ the average of the leading entropy term $\overline{{\cal S}(\{k_i\})}$ is well defined, with vanishing relative error.\\

Since the leading entropy term ${\cal S}(\{k_i\})=\frac{1}{2}\avg{k}N \ln N$ depends only on the average degree $\avg{k}$, we can further analyze entropy fluctuations in the ensemble with a fixed $\avg{k}$. Therefore we next evaluate the conditional probability distribution $P(S|\avg{k})$ that depends only on the distribution of the subleading entropy term, since average degree $\avg{k}$ determines uniquely the leading term.

\subsection{Conditional entropy distribution $P(S|\avg{k})$}

If $P(\{k_i\})$ is the probability of degree sequence $\{k_i\}$, then
\bea
P(S|\avg{k})&=&\int \prod_i dk_i P(\{k_i\})\delta\left(S,S(\{k_i\})\right)\nonumber\\
&\times&\delta\left(\avg{k}N,\sum_{i=1}^Nk_i\right).
\label{ps00}
\eea
Since entropy $S(\{k_i\})=\frac{1}{2}\avg{k}N\ln N-N\sigma(\{k_i\})$ is a function of hidden variables $\{h_i\}$ only, $S(\{k_i\})=S(\{h_i\})$,
we can perform the following change of variables in the last equation:
\bea
P(\{k_i\})\prod_{i=1}^N dk_i=\Pi(\{h_i\})\prod_{i=1}N dh_i,
\eea
where $\Pi(\{h_i\})$ is the probability of hidden variables sequence $\{h_i\}$.
After this transformation, and expressing delta functions in  Eq.~(\ref{ps00}) via exponentials, we obtain,
\bea
P(S|\avg{k})&=&\int \prod_i dh_i \Pi(\{h_i\})\int d\omega e^{i\omega [S-S(\{h_i\})]}\times\nonumber \\
&&\times\int d\nu e^{i\nu\left[\avg{k}N-\sum_{i,j| i\neq j}p(h_i,h_j)\right]}
\label{ps0}
\eea
We next make the simplifying assumption  that the hidden variables are i.i.d.\ distributed, $\Pi(\{h_i\})=\prod_{i=1}^N \tilde{\pi}(h_i)$ with some distribution $\tilde{\pi}(h_i)$.
In the large network limit we can then transform the multiplex integral over $N$ variables $\{h_1,h_2,\ldots, h_N\}$ to a functional integral over density function
\bea
\rho(h)=\frac{1}{N}\sum_{i=1}^N \delta(h,h_i),
\eea
imposing constraint $\int dh \, \rho(h)=1$ by Lagrangian multiplier $\mu$. The distribution $P(S|\avg{k})$ defined in Eq.~(\ref{ps0}) becomes
\bea
\hspace*{-3mm}P(S|\avg{k})=\int d\omega \int d\mu \int d\nu \int {\cal D}\rho(h) e^{G(\rho,\mu,\omega,\nu)},
\label{ps1}
\eea
 with
 \bea
&& G(\rho,\mu, \omega,\nu)=-N\int dh \rho(h) \ln\left[\frac{\rho(h)}{\tilde{\pi}(h)}\right]\nonumber \\
 &&-i N^2\int dh \int dh' \rho(h) \rho(h') [\omega s(h,h')+\nu p(h,h')]\nonumber \\
&& -i\mu N\int dh \left[\rho(h)-1\right]+i\nu \avg{k}N+i\omega S,
 \label{G}
 \eea
  where
\bea
s(h,h')&=&-\frac{1}{2}\left\{\frac{hh'/N}{1+hh'/N}\ln\left[\frac{hh'/N}{1+h h'/N}\right]\right.\nonumber \\
&&\left.+\frac{1}{1+hh'/N}\ln\left[\frac{1}{1+h h'/N}\right]\right\}\nonumber \\
p(h,h')&=&\frac{hh'/N}{1+hh'/N}.
\label{shh'phh'}
\eea
The integrals in Eq.~(\ref{ps1}) can be evaluated at the saddle point given by
\bea
S&=&N^2\int dh \int dh' \rho(h) \rho(h') \sigma(h,h')\label{sp} \\
\avg{k}&=&N^2\int dh \int dh' \rho(h) \rho(h') p(h,h') \nonumber \\
\rho(h)&=&\frac{\tilde{\pi}(h) e^{-2N \int dh' \rho(h') [\omega s(h,h')+\nu p(h,h')]}}{\int dh'' \tilde{\pi}(h'') e^{- 2N \int dh' \rho(h') [\omega s(h'',h')+\nu p(h'',h')]}},\nonumber
\eea
where we have performed the Wick rotation of parameters $\omega $ and $\nu$.
Denoting by $\rho^{\star}(h)$ the $S$-dependent solution of the above saddle point equations,
we obtain the following simple expression for distribution $P(S|\avg{k})$:
\bea
P(S|\avg{k})&=& e^{-ND_{KL}\left[\rho^{\star}(h)|\tilde{\pi}(h)\right]},\,\,\text{where}\\
D_{KL}\left[\rho^{\star}(h)|\tilde{\pi}(h)\right] &=& \int dh\,\rho^{\star}(h)\ln\frac{\rho^{\star}(h)}{\tilde{\pi}(h)}
\eea
is the Kullback-Leibler distance between distributions $\rho^{\star}(h)$ and $\tilde{\pi}(h)$.
Therefore conditional distribution $P(S|\avg{k})$ is well behaved, and depends only on KL-distance $D_{KL}\left[\rho^{\star}(h)|\tilde{\pi}(h)\right]$.

\subsection{Condensation as a large deviation event}

\begin{figure}
\begin{center}
{\includegraphics[width=3.1in]{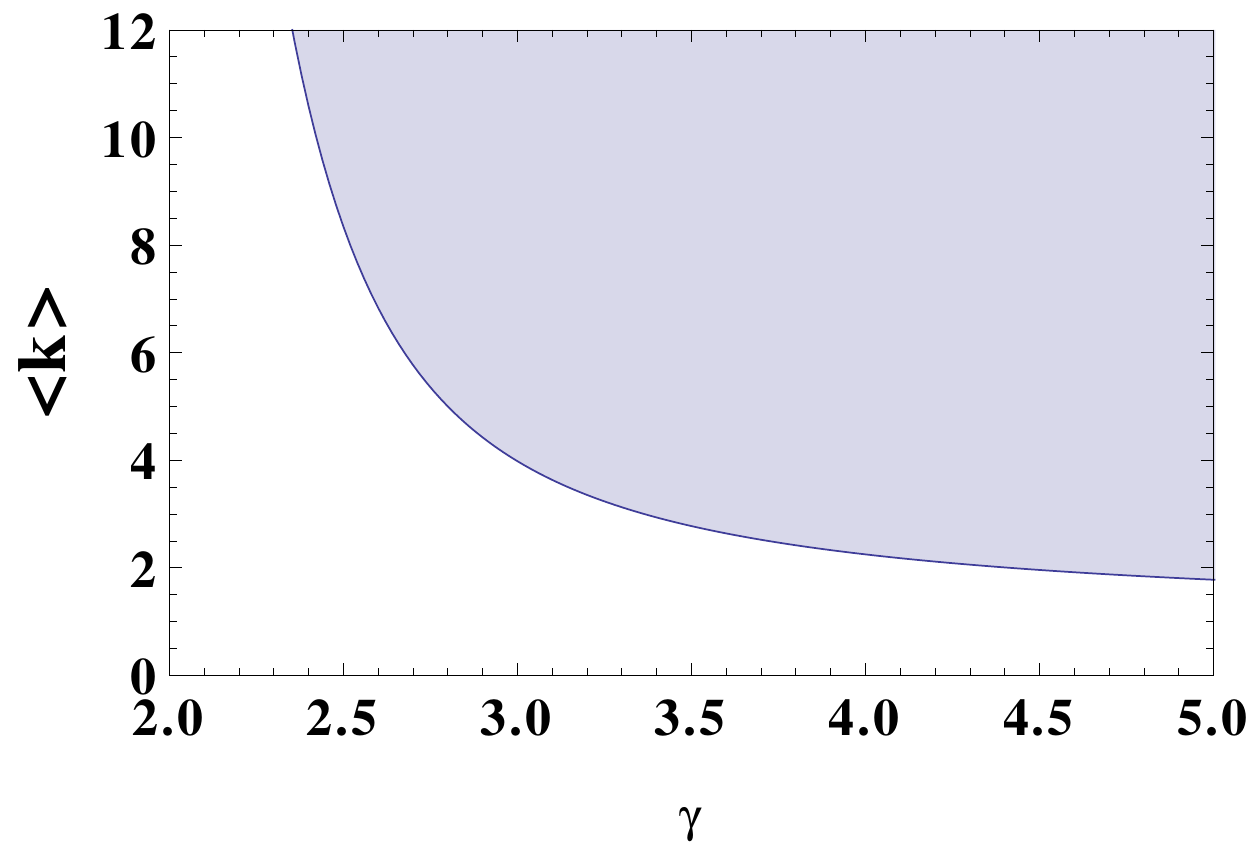}}
\end{center}
\caption{Phase diagram showing the region of the plain $(\gamma,\avg{k})$ where we observe the condensation discussed in the text (shaded region). The figure corresponds to the hidden variable distribution $\tilde{\pi}(h)\propto h^{-\gamma}$ with $h\in[1,N]$ and $N=10^4$.}
\label{fig1}
\end{figure}

The saddle point Eqs.~(\ref{sp}) have a solution only if the average degree $\avg{k}$ is equal to or less than the expected degree $m$ of the degree distribution, i.e. only if
 \bea
 \avg{k}\leq m=N\int dh \int dh' \, \tilde{\pi}(h')\tilde{\pi}(h) p(h,h').
 \eea
In fact the Lagrangian multipliers $\omega,\nu$ must be real and greater than zero to guarantee that $\rho(h)$ given by Eq.~(\ref{sp}) is well defined.
Following \cite{Marsili}, to explore large deviation properties of a fat-tailed distribution, we use the following ansatz:
 \bea
 N\rho(h)=(N-1)\rho_c(h)+\delta(h,h_c).
 \eea
This ansatz accounts for a spontaneous breaking of permutation symmetry between the $N$ hidden variables in the ensemble, and reflects our expectation to detect condensation in some large-deviation realization in the ensemble.
With this ansatz, probability $P(S|\avg{k})$ becomes
 \bea
\hspace*{-3mm}P(S|\avg{k})=\int d\omega \int d\mu \int d\nu \int {\cal D}\rho(h) e^{G(\rho_c,h_c,\mu,\omega,\nu)},
\label{ps1b}
\eea
 with
 \bea
&& G(\rho_c,h_c,\mu, \omega,\nu)=-N\int dh \rho_c(h) \ln\left[\frac{\rho_c(h)}{\tilde{\pi}(h)}\right]\nonumber \\
&&-\frac{1}{N}\ln \tilde{\pi(h_c)}\nonumber \\
 &&-i N^2\int dh \int dh' \rho_c(h) \rho_c(h') [\omega s(h,h')+\nu p(h,h')]\nonumber \\
 &&-i 2 N^2 \int dh \rho_c(h)  [\omega s(h_c,h)+\nu p(h_c,h])\nonumber \\
&& -i\mu N \int dh \, \left[\rho_c(h)-1\right]+i\nu\avg{k}N+i\omega S,
 \label{Gb}
 \eea
where functions $s(h,h')$ and $p(h,h')$ are as in Eqs.~(\ref{shh'phh'}).
The problem of minimizing function $G(\rho_c,h_c,\mu,\omega,\nu)$ with respect to all its parameters has a non-trivial solution only if $\avg{k}>m$, in  which case we have
\bea
\avg{k}N=mN+2N^2\int dh  \rho_c(h)   p(h_c,h)
\eea
with $\rho_c^{\star}(h)=\tilde{\pi} (h)$, and
$G(\rho_c,h_c,\mu,\omega,\nu)=0$ for any $\avg{k}>m$.
In Figure~$\ref{fig1}$ we show the phase diagram $(\gamma,\avg{k})$, where $\gamma$ is the exponent of the hidden variable distribution $\tilde{\pi}(h)\propto h^{-\gamma}$, and $h\in [1,N]$ for $N=10^4$. Above the curve $\avg{k}=m$, i.e., in the shaded region of parameter values, we observe condensation.

\section{Entropy distribution in the network ensemble with a given distribution of exact degrees}

The results presented in the previous section concerning the soft ensembles remain qualitatively unchanged if we consider the hard ensembles of networks with a given degree distribution of exact degrees. Note that we are treating here always quenched disorder. In fact here we consider ensembles of networks of fixed degree sequence, where each degree sequence is drawn randomly from a given degree distribution. As in the soft case, in this hard case we assume that the disorder is quenched, and that the exact degree sequence $\{k_i\}$ is fixed but unknown and drawn from degree distribution $p(k)$.
The probability of degree sequence $\{k_i\}$ is thus $P(\{k_i\})=\prod_i p(k_i)$, and for each $\{k_i\}$
we consider the microcanonical ensemble of networks with the fixed sequence of exact degrees $\{k_i\}$. In this ensemble, network $G$ has probability  $P_M(G|\{k_i\})$ defined in Eq.~(\ref{PM}), so that the probability distribution $P_M(G)$ in the ensemble is
\bea
P_M(G)=\int {\prod_{i=1}^Nd{k_i}}   P(\{k_i\}) P_M(G|\{k_i\}).
\eea
Given degree sequence $\{k_i\}$, and using  Eq.~(\ref{SSO}) and Eq.~(\ref{S2}), the Gibbs entropy $N\Sigma(\{k_i\})$ is the sum of three contributions,
\bea
N\Sigma(\{k_i\})&=&S(\{k_i\})-N\Omega(\{k_i\})\nonumber \\&=&{\cal S}(\{k_i\})-N\sigma(\{k_i\})-N\Omega(\{k_i\},
\eea
where the leading term of $N\Sigma(\{k_i\})$ is ${\cal S}(\{k_i\})=\frac{1}{2}\avg{k}N\ln N$.
In what follows we analyze the entropy distribution $P(N\Sigma)$ in the ensemble
\bea
P(N\Sigma)=\int {\prod_{i=1}^N dk_i}   P(\{k_i\}) \delta\left[N\Sigma,N\Sigma(\{k_i\})\right].
\eea
If we cannot compute the full distribution $P(N\Sigma)$ exactly, we may still characterize
its average, variance, and relative error
\bea
\overline{N\Sigma(\{k_i\})}&=& \int {\prod_{i=1}^N dk_i}   P(\{k_i\}) N\Sigma(\{k_i\})\nonumber \\
\overline{\left[\delta N\Sigma(\{k_i\})\right]^2}&=& \int \prod_{i=1}^Nd{k_i}  P(\{k_i\})
 \left[N\Sigma(\{k_i\})-\overline{N\Sigma(\{k_i\})}\right]^2,\nonumber \\
 \Delta [N\Sigma(\{k_i\})]&=&\frac{\sqrt{\overline{\left[\delta N\Sigma(\{k_i\})\right]^2}}}{\overline{N\Sigma(\{k_i\})}}.\nonumber
\eea

\subsection{Entropy distribution in scale-free networks}
We first consider the probability distribution $P({\cal S})$ of the leading term ${\cal S}$ of entropy $N\Sigma(\{k_i\})$, defined as
\bea
P({\cal S})=\int {\prod_{i=1}^N dk_i}  \prod_{i=1}^N p(k_i) \delta\left[{\cal S},{\cal S}(\{k_i\})\right],
\eea
where  ${\cal S}(\{k_i\})=\frac{1}{2}\avg{k} N\ln N$ depends only on the average degree in the network.
According to the Generalized Central Limit theorem \cite{Bouchaud}, and similarly to the soft case,  we have the following two cases:
\begin{itemize}
\item {\em Case $\gamma>3$}:\\
The distribution $P({\cal S})$ converges to a Gaussian distribution and the relative error on the average of ${\cal S}$ is vanishing in the large network limit. In fact we find that
\bea
 \Delta {\cal S}(\{k_i\})\propto N^{-1/2}.
\eea
\item {\em Case $\gamma\in (2,3]$}:\\
Due to the structural degree cutoff $k_i\leq k_{\max}=N$, the entropy distribution has a vanishing relative error given by
\bea
 \Delta {\cal S}(\{k_i\})\propto N^{-(\gamma-2)/2}.
\eea

\end{itemize}
In the both  cases, the average $\overline{{\cal S}(\{k_i\})}$ is well defined with a  relative error $\Delta {\cal S}(\{k_i\})$ vanishing in the large network limit.

\subsection{Conditional entropy distribution $P(N\Sigma|\avg{k})$}
Similarly to the soft case, we next show that the large entropy fluctuations are due exclusively to the fluctuations of the total number of links in the ensemble.
We note that these fluctuations are necessarily present since the exact degree sequence in each network in the ensemble is independently sampled from the given distribution.
Following a similar procedure, we evaluate the probability of $N\Sigma$ conditioned to a fixed  value of the average degree in the network $\avg{k}$, $P(N\Sigma|\avg{k})$. This conditional entropy distribution depends only on the distribution of the subleading contributions to $N\Sigma(\{k_i\})$, $N\sigma(\{k_i\})+N\Omega(\{k_i\})$, because the average degree determines uniquely the leading term ${\cal S}(\{k_i\})$.

If $P(\{k_i\})$ is the probability of degree sequence $\{k_i\}$, then
\bea
P(N\Sigma|\avg{k})&=&\int \prod_i dk_i P(\{k_i\})\delta\left(N\Sigma,N\Sigma(\{k_i\}\right)\nonumber\\
&\times&\delta\left(\avg{k}N,\sum_{i=1}^Nk_i\right).
\label{ps00b}
\eea
Since entropy $S(\{k_i\})=\frac{1}{2}\avg{k}N\ln N-N\sigma(\{k_i\})+N\Omega(\{k_i\})$ is a   function of hidden variables $\{h_i\}$ only,
$N\Sigma(\{k_i\})=N\Sigma(\{h_i\})$, we can change variables
\bea
P(\{k_i\})\prod_{i=1}^n dk_i=\Pi(\{h_i\})\prod_{i=1}^N dh_i
\eea
where $\Pi(\{h_i\})$ is the probability of hidden variables sequence $\{h_i\}$, and obtain,
\bea
\hspace*{-1mm}P(N\Sigma|\avg{k})&=&\int \prod_i dh_i \Pi(\{h_i\})\int d\omega e^{i\omega [N\Sigma-N\Sigma(\{h_i\})]}\times\nonumber \\
&&\times\int d\nu e^{i\nu\left[\avg{k}N-\sum_{i,j| i\neq j}p(h_i,h_j)\right]}.
\label{ps0b}
\eea
Assuming next that our hidden variables are i.i.d.\ distributed $\Pi(\{h_i\})=\prod_{i=1}^N \tilde{\pi}(h_i)$ with some distribution $\tilde{\pi}(h_i)$,
we transform the multiplex integral over $N$ variables $\{h_1,h_2,\ldots, h_N\}$ in the large network limit to a functional integral over density function
\bea
\rho(h)=\frac{1}{N}\sum_{i=1}^N \delta(h,h_i),
\eea
imposing constraint $\int dh \rho(h)=1$ by Lagrangian multiplier $\mu$. The distribution $P(N\Sigma|\avg{k})$ defined in Eq.~(\ref{ps0b}) becomes
\bea
\hspace*{-3mm}P(N\Sigma|\avg{k})=\int d\omega \int d\mu \int d\nu \int {\cal D}\rho(h) e^{G(\rho,\mu,\omega,\nu)},
\label{ps1bb}
\eea
 with
 \bea
&& G(\rho,\mu, \omega,\nu)=-N\int dh \rho(h) \ln\left[\frac{\rho(h)}{\tilde{\pi}(h)}\right]\nonumber \\
 &&-i N^2\int dh \int dh' \rho(h) \rho(h') [\omega s(h,h')+\nu p(h,h')]\nonumber \\
 &&+i \omega N \int dh \rho(h)\ln \left(\frac{k(h)^{k(h)}e^{-k(h)}}{k(h)!} \right)\nonumber \\
&& -i\mu N\int dh\left[ \rho(h)-1\right]+i\nu \avg{k}N+i\omega N\Sigma,
 \label{Gb2}
 \eea
 where \bea
s(h,h')&=&-\frac{1}{2}\left\{\frac{hh'/N}{1+hh'/N}\ln\left[\frac{hh'/N}{1+h h'/N}\right]\right.\nonumber \\
&&\left.+\frac{1}{1+hh'/N}\ln\left[\frac{1}{1+h h'/N}\right]\right\}\nonumber \\
p(h,h')&=&\frac{hh'/N}{1+hh'/N},\nonumber \\
k(h)&=&N\int dh' \rho(h') p(h,h').
\label{shh'phh'b}
\eea
The integrals in Eq.~(\ref{ps1bb}) can be evaluated at the saddle point given by
\bea
N\Sigma &=&N^2\int dh \int dh' \rho(h) \rho(h') \sigma(h,h')\nonumber \\
&&- N\int dh \rho(h)\ln \left(\frac{k(h)^{k(h)}e^{-k(h)}}{k(h)!}\right) \\
k(h)&=&N\int dh' \rho(h') p(h,h')\nonumber \\
\avg{k}&=&N^2\int dh \int dh' \rho(h) \rho(h') p(h,h') \nonumber \\
\rho(h)&=&\frac{1}{{\cal C}}\tilde{\pi}(h)\left(\frac{k(h)^{k(h)}e^{-k(h)}}{k(h)!}\right)^{\omega} \nonumber \\
&&\hspace{-10mm}\times \exp\left\{-N \int dh' \rho(h')p(h,h')\left[2\nu+\omega (H_{k(h')}-\ln k(h'))\right]\right\}\nonumber \\
&&\times \exp\left\{-N \int dh' \rho(h') 2\omega s(h,h')\right\}\label{spb}
\eea
where ${\cal C}$ is a normalization constant and $H_{k(h)}$ stands for the Harmonic number.
Denoting by $\rho^{\star}(h)$ the $N\Sigma$-dependent solution of the above saddle point equations, we get the following simple expression for
distribution $P(N\Sigma|\avg{k})$:
\bea
P(N\Sigma|\avg{k})&=& e^{-ND_{KL}\left[\rho^{\star}(h)|\tilde{\pi}(h)\right]}.
\eea
As in the soft case, in this hard case the entropy distribution conditioned on the average degree in the network is well-behaved and depends only on the KL distance $D_{KL}\left[\rho^{\star}(h)|\tilde{\pi}(h)\right]$ between distributions $\rho^{\star}(h)$ and $\tilde{\pi}(h)$.

 \subsection{Condensation as a large deviation event}

 The saddle point Eqs.~(\ref{spb}) have a solution only if  the average degree $\avg{k}$ is equal to or less than the expected degree $m$ over the degree distribution,\bea
 \avg{k}<m=N\int dh \int dh' \tilde{\pi}(h')\tilde{\pi}(h) p(h,h').
 \eea
 In fact the  Lagrangian multipliers $\omega,\nu$ must be real and greater than zero to guarantee that $\rho(h)$ given by Eq.~(\ref{spb}) is well defined.
 Therefore, following the same logic as in the soft case,  we use the following ansatz:
 \bea
 N\rho(h)=(N-1)\rho_c(h)+\delta(h,h_c),
 \eea
With this this ansatz, probability $P(N\Sigma|\avg{k})$ becomes
 \bea
\hspace*{-3mm}P(N\Sigma|\avg{k})=\int d\omega \int d\mu \int d\nu \int {\cal D}\rho(h) e^{G(\rho_c,h_c,\mu,\omega,\nu)},
\eea
 with
 \bea
&& G(\rho_c,h_c,\mu, \omega,\nu)=-N\int dh \rho_c(h) \ln\left[\frac{\rho_c(h)}{\tilde{\pi}(h)}\right]\nonumber \\
&&-\frac{1}{N}\ln \tilde{\pi(h_c)}\nonumber \\
 &&-i N^2\int dh \int dh' \rho_c(h) \rho_c(h') [\omega s(h,h')+\nu p(h,h')]\nonumber \\
 &&+i \omega N \int dh \rho_c(h)\ln \left(\frac{k(h)^{k(h)}e^{-k(h)}}{k(h)!} \right)\nonumber \\
 &&-i 2 N^2 \int dh \rho_c(h)  [\omega s(h_c,h)+\nu p(h_c,h])\nonumber \\
&& +i \omega N \ln \left(\frac{k(h)^{k(h_c)}e^{-k(h_c)}}{k(h_c)!} \right)\nonumber \\
&& -i\mu N\int dh \left[\rho(h)-1\right]+i\nu \avg{k}N+i\omega N\Sigma,
 \label{Gbb}
 \eea
where functions $s(h,h')$, $p(h,h')$,  and $k(h)$ are as in Eqs.~(\ref{shh'phh'b}).
The problem of minimizing function $G(\rho_c,h_c,\mu,\omega,\nu)$ with respect to all its parameters, has a non-trivial solution only if $\avg{k}>m$, in which case we have
\bea
\avg{k}N=mN+2N^2\int dh  \rho_c(h)   p(h_c,h)
\eea
with $\rho_c^{\star}(h)=\tilde{\pi} (h)$,
and $G(\rho_c,h_c,\mu,\omega,\nu)=0$ for any $\avg{k}>m$.
The phase diagram of $(\gamma,\avg{k})$ with $\tilde{\pi}(h)\propto h^{-\gamma}$ and $h\in (1,N)$ for $N=10^4$ is the same as in the soft case shown in Figure~\ref{fig1}.
If $\avg{k}>m$, we can observe condensation---a single node in the network can acquire a degree of the order of $N$.

\section{Conclusion}

Motivated by the observation that in modeling real networks, the degree distribution is a more reasonable and realistic constraint than the degree sequence, we have studied the entropy distribution in the ensembles of random networks with a given degree distribution.
We found that entropy is self-averaging, thanks to the structural degree cutoff at $k_{\max}=N$. The fluctuations of entropy are mainly determined by the fluctuations of the average degree in the ensembles.  Networks with average degree exceeding a certain threshold, $\avg{k}>m$, exhibit  large deviation or condensation effects---a single node can attract $O(N)$ links. Interestingly, this condensation is different from the Bose-Einstein condensation in complex networks \cite{BE} in that the condensation considered here corresponds only to some ``large deviation configurations.'' It is not typically expected in the ensemble.

{\it Disclaimer}: This work does not reflect the official view of the Bank of Canada.

\acknowledgments
We thank M.~Marsili and M.~Ostilli for interesting discussions. This work was supported by NSF grants No.\ CNS-1344289, CNS-1039646, and CNS-0964236; DARPA grant No.\ HR0011-12-1-0012; and by Cisco Systems.

\end{document}